\documentclass[amssymb,aps,pre,twocolumn,showpacs,superscriptaddress,groupedaddress]{revtex4-1}  

\usepackage{graphicx}  
\usepackage{dcolumn}   
\usepackage{bm}        
\usepackage{amssymb}   
\usepackage{appendix}
\usepackage{color}

\DeclareMathAlphabet{\mabold}{OT1}{cmr}{bx}{it}

\begin{document}

\title{Role of Thermal Fluctuations in Nonlinear Thin Film Dewetting}

\author{S.\ Nesic,$^1$ R.\ Cuerno,$^1$ E.\ Moro$^{1}$, L. Kondic$^3$}
\affiliation{$^1$Departamento de Matem\'aticas \& Grupo Interdisciplinar de Sistemas Complejos (GISC), Universidad Carlos III de Madrid, 28911 Legan\'es, Spain}
\affiliation{$^3$Department of Mathematical Sciences, New Jersey Institute of Technology, Newark, New Jersey}

\date{\today}

\begin{abstract}
The spontaneous formation of droplets via dewetting of a thin fluid film from a solid substrate allows for materials nanostructuring, under appropriate experimental control. While thermal fluctuations are expected to play a role in this process, their relevance has remained poorly understood, particularly during the nonlinear stages of evolution. Within a stochastic lubrication framework, we show that thermal noise speeds up and substantially influences the formation and evolution of the droplet arrangement. As compared with their deterministic counterparts, for a fixed spatial domain, stochastic systems feature a smaller number of droplets, with a larger variability in sizes and space distribution. Finally, we discuss the influence of stochasticity on droplet coarsening for very long times.
\end{abstract}

\maketitle

The interplay between stochastic fluctuations and nonlinear interactions can induce highly nontrivial effects in spatially extended systems \cite{sagues2007} at the nanoscale \cite{landman2005}. For instance, noise can rectify the direction of material transport, as for diffusing particles under asymmetric forces \cite{hanggi2009}. When a characteristic pattern emerges from a homogenous state \cite{cross}, fluctuations can even enhance (rather than hinder) spatial order, or modify the rate at which typical pattern sizes increase with time ({\em coarsening}), as e.g.\ for evolving atomic steps at epitaxial surfaces \cite{politi2000,misbah2010}.

A natural context in which fluctuations are expected to be relevant is nanoscale fluid flow. Although the continuum framework provided by the Navier-Stokes equations is physically valid down to surprisingly small scales ($\simeq 1$ nm) \cite{bocquet2010,detcheverry2013}, the atomistic nature of the fluid medium is expected to play an increasingly important role as physical scales are reduced. Note that in this process the surface-to-volume ratio also becomes ever more favorable \cite{allara2005}, so that free surface flows \cite{craster2009} provide conspicuous instances for noise effects. Thus, a number of interfacial processes have been seen to depend critically on the occurrence of fluctuations, such as e.g.\ the breaking of nanojets \cite{moseler2000,eggers2002,hennequin2006} or of liquid threads \cite{petit2012}. In addition, it is known that thermal noise in the fluid velocity field changes the value of the contact angle under partial wetting conditions \cite{nesic2015}, and enhances the spreading of droplets in surface-tension \cite{davidovitch2005} and gravity \cite{nesic2015} driven systems, as well as the rupture of thin dewetting films \cite{becker2003,grun2006}.

Indeed, in dewetting experiments carried out using polymer \cite{fetzer2007} or liquid metal films \cite{gonzalez2013}, dynamics and morphologies are not reproducible by deterministic frameworks. Early rupture times and irregular patterns suggest that fluctuations play a strong dynamical role. Working in the long wave (lubrication) approximation to free surface flow \cite{oron1997}, previous works improving deterministic models \cite{mitlin2000,becker2003,davidovitch2005,grun2006,fetzer2007} cast the time evolution of the system in the form of a stochastic evolution equation (SLE) for the thickness $h$ of the thin fluid film, namely,
\begin{equation}
\eta \partial_t h =  \partial_x\, \left\{\frac{h^3}{3}\,\partial_x\left[-\gamma \partial_x^2 h - \Pi(h) \right]+ \sigma\, h^{3/2}\,\epsilon(x,t)\right\},
 \label{stochastic_thin_films}
\end{equation}
where $\eta$ is viscosity, $\gamma$ is surface tension, $\epsilon(x,t)$ is a Gaussian white noise of zero mean and unit variance, and
$\sigma = \sqrt{\eta k_B T/3}$ depends on temperature $T$. In Eq.\ (\ref{stochastic_thin_films}), $\Pi(h)=-\partial \Phi(h)/\partial h$ is the disjoining pressure that accounts for fluid-solid interaction, with $\Phi(h)$ the interface potential \cite{butt}. A power law is commonly used, $\Pi(h)=\kappa[ (h_*/h)^n-(h_*/h)^m]$, where $\kappa$ is proportional to the Hamaker constant and $h_*$ is the
precursor film thickness \protect{\cite{mitlin2000,bonn2009}}, corresponding to the minimum of the potential.  The equilibrium contact angle emerges from the interplay between surface tension and disjoining pressure, see e.g.~\protect{\cite{diez2007}}.

As shown in \cite{grun2006,fetzer2007}, for short-times Eq.\ (\ref{stochastic_thin_films}) predicts a morphological instability \cite{cross}: surface tension, in competition with the destabilizing disjoining pressure, selects a typical length scale, $\lambda$, of surface undulations. In a process reminiscent of domain coarsening in phase separation or spinodal decomposition of binary mixtures \cite{bray1994}, and as seen in the experiments \cite{fetzer2007}, this scale increases nontrivially with time in a form which cannot be accounted for deterministically [i.e., for $\sigma=0$ in Eq.\ (\ref{stochastic_thin_films})]. However, the study in~\protect{\cite{fetzer2007}} was limited to the linear regime, where perturbations of the flat film are small compared to its thickness and the morphology is dominated by capillary-like surface modes.
Pressing open questions then are if and how do the nonlinearities, responsible for actual droplet formation in Eq.\ (\ref{stochastic_thin_films}), modify this time evolution. Moreover, at very long times deterministic droplets
are known to undergo coarsening into a single-drop morphology \cite{glasner2003,limary2003}. Hence, one can ponder whether thermal fluctuations modify the coarsening behavior in this case. For instance, noise is known to modify the coarsening law of the 1D Cahn-Hilliard equation, a paradigmatic model in the context of spinodal decomposition \cite{bray1994,torcini2002}.

In this Letter we study the effect of thermal fluctuations on the formation and evolution of droplets under partial wetting conditions. To this end, we study numerically the SLE [Eq.~(\ref{stochastic_thin_films})] in the nonlinear regime. We find that thermal fluctuations unambiguously speed up the nonlinear process of droplet formation. Moreover, we show that, as compared to the deterministic case, noise increases heterogeneity in droplet sizes and positions, while seemingly not affecting the coarsening process expected for asymptotically long times \cite{glasner2003,limary2003}.

To study the behavior of the solutions of Eq.~(\ref{stochastic_thin_films}) in nonlinear regime, we have carried out
large-scale numerical simulations using a scheme \cite{diez2000} that ensures non-negativeness of the solution for all times, if the initial film is positive in the full domain \cite{zhornitskaya2000}. Specifically, our algorithm is based on the standard implicit (Crank-Nicholson) discretization~\cite{diez2000}, where surface tension is treated implicitly, while $\Pi(h)$ is treated explicitly; we employ zero-flux boundary conditions. The stochastic term in Eq.\ (\ref{stochastic_thin_films}) is also dealt with explicitly, within the Stratonovich interpretation \cite{grun2006}.

In our simulations we consider a nondimensional version of Eq.~(\protect{\ref{stochastic_thin_films}}) obtained by defining
$\hat h=h/h_c$, $\hat x = x/h_c$, and $\hat t=t/t_c$, where $h_c$ is a typical film thickness and $t_c=3\eta h_c/\gamma$ \cite{SM_sq}. This leads to non-dimensional amplitudes $\hat\sigma=(k_B T/\gamma h_c^2)^{1/2}$ and to $\hat{\kappa}=\kappa h_c/\gamma$; we use the exponents $(n,m)=(3,2)$ as in e.g.~\cite{diez2007}.
We perform the deterministic and stochastic simulations of Eq.~(\protect{\ref{stochastic_thin_films}}) using a precursor thickness $\hat h_*=0.01$ and the same random {\em initial condition}, namely, random values of the thickness with non-dimensional average $\hat h_0=0.1$ and variance $10^{-2} \hat h_0$.
The contact angle is set to $50^\circ$ in the expression $\hat\kappa = 2 (1-\cos\theta) /\hat{h}_*$  \cite{diez2007}, leading to $\hat \kappa = 72$; within the long wave theory implementation, the actual contact angle (measured by the slope of the tangent line passed through the drop profile through the point
of inflection) is smaller and is close to $25^\circ$.
The spatial step size $dx=\hat h_*$; this choice is known to lead to accurate results~\protect{\cite{diez2000}}. The temporal
step size is adaptive, following the approach described in~\protect{\cite{diez2000}}. This
allows to obtain converged results with reasonable computational effort; the use of such adaptive time stepping is particularly important for the purpose of carrying out simulations for long times where coarsening effects become relevant. We use a large domain
size $L\approx 31 \lambda$, where $\lambda$ is the most unstable wavelength obtained by linear stability analysis of deterministic
version of Eq.~(\ref{stochastic_thin_films}), discussed further below. Combined with a large number of realizations, $\simeq 200$, such a domain allows to obtain statistically meaningful results.
The specified  parameter values are closely related to the polymer films studied in \cite{fetzer2007}, where the characteristic film thickness is $4$ nm ($h_c=40$ nm, so that $\hat h_0=0.1$), while $\gamma=0.03$ N/m and the Hamaker constant $A=2\cdot 10^{-20}$ J yield a contact angle in the $15-20^\circ$ range. For these parameter values the non-dimensional noise strength $\hat\sigma\simeq10^{-2}$ corresponds to $T=50-60^\circ$ C. On the other hand, for the liquid metal thin films considered in \cite{gonzalez2013}, $\gamma=1.3$ N/m, $T=2000$ K, and $h_c\in [50,150]$ nm, leading to $\hat\sigma\in [10^{-5/2},10^{-3}]$.

\begin{figure}[t!]
\includegraphics[width=8cm]{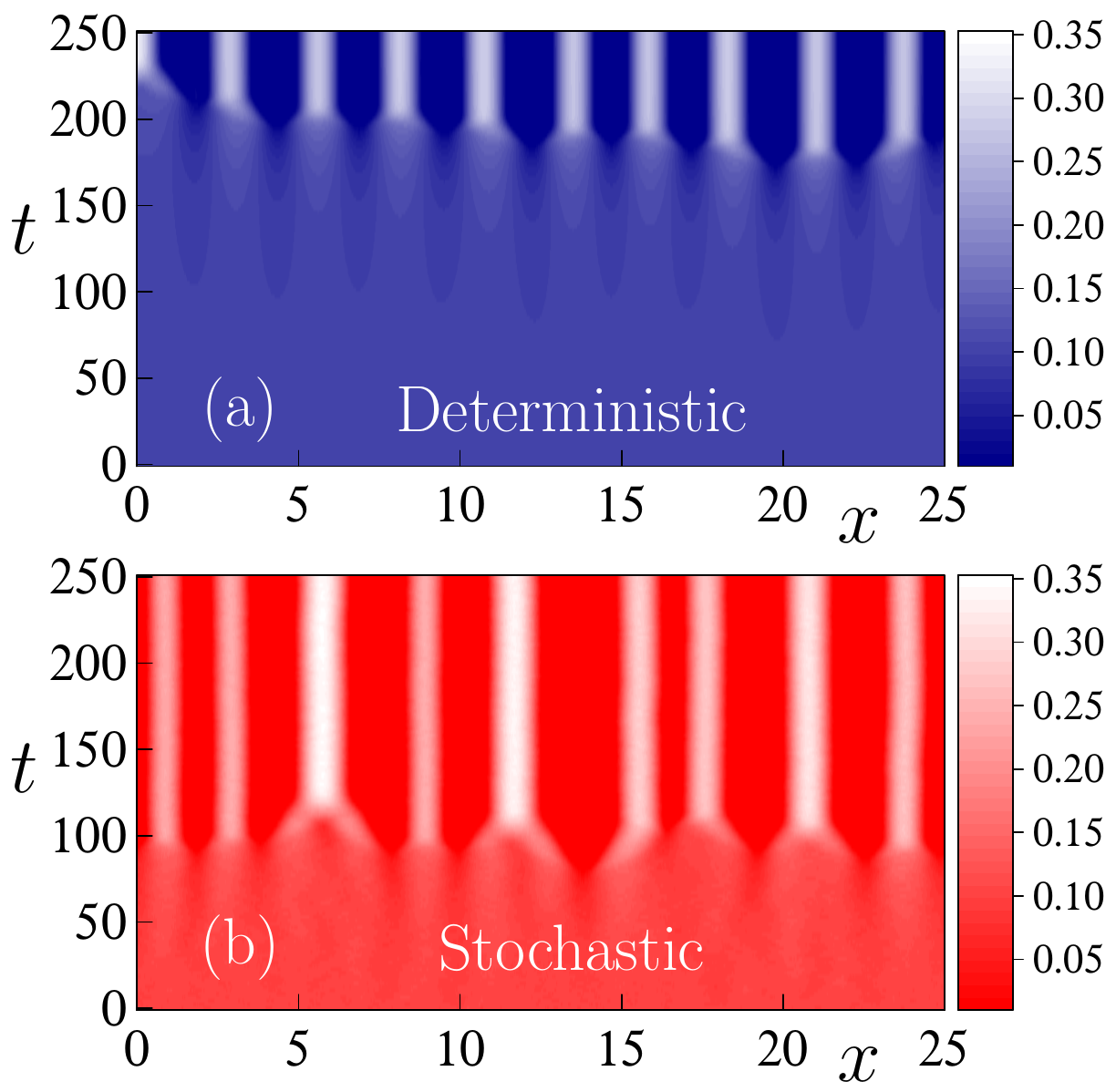}
\caption{\label{fig:surface_in_time} Space-time plot of droplet formation and evolution as predicted by Eq.\ (\ref{stochastic_thin_films}) for $\sigma=0$ (a) and $\sigma=10^{-2}$ (b), for the same parameter values and initial conditions, see main text.
Brighter (darker) color corresponds to larger (smaller) values of the film thickness $h(x,t)$.
}
\end{figure}

From now on, and unless otherwise stated, we work in dimensionless units and remove hats for notational simplicity. Figure \ref{fig:surface_in_time} shows  examples of the time evolution predicted by Eq.\ (\ref{stochastic_thin_films}) in the deterministic and stochastic cases.
Well-defined droplets (clear bands) emerge after a {\em rupture time} of roughly $t_{\rm r, det}=180$ ($t_{\rm r,sto}=80$) time units in the deterministic (stochastic) system. In the latter case there is a substantial amount of droplet merging around that time, after which activity decreases. Comparing both panels, we immediately observe that the width of the droplets (clear bands) and their locations are much more regular in the deterministic than in the stochastic case.

\begin{figure}[t!]
\includegraphics[width=8.5cm]{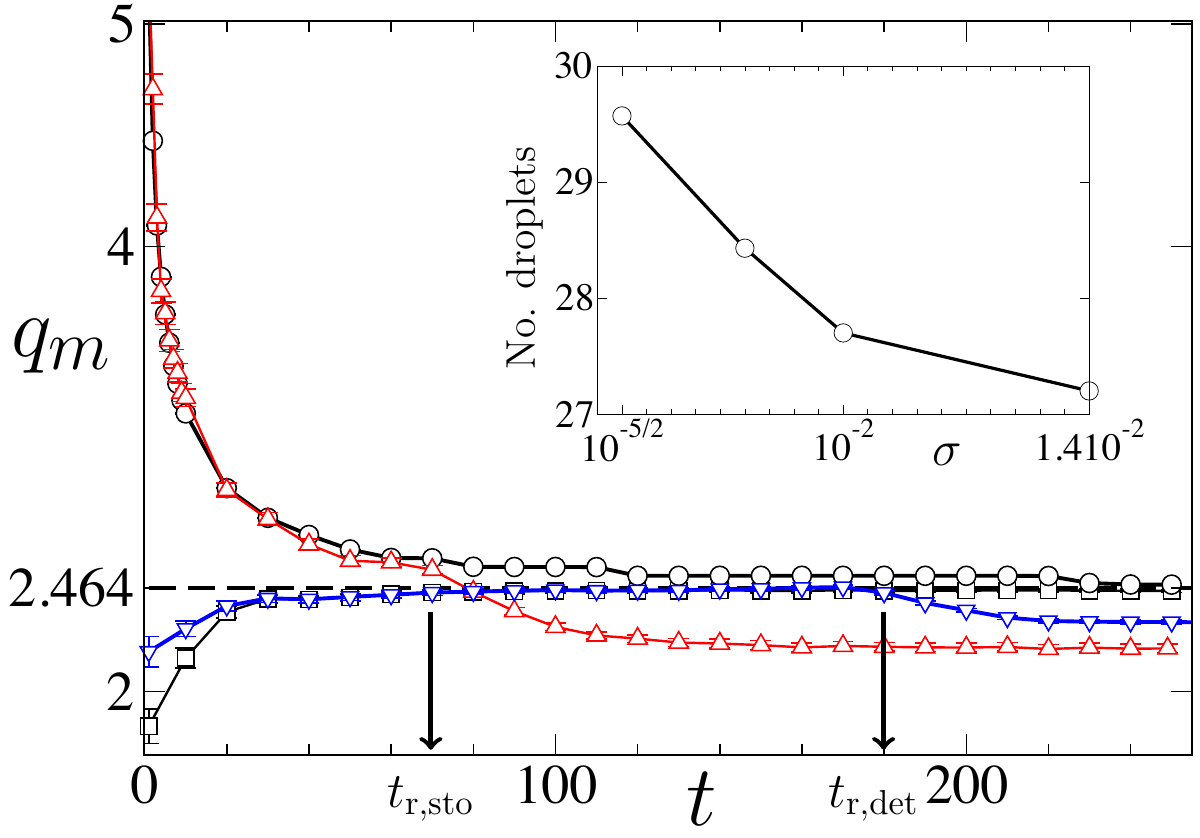}
\caption{\label{fig:peak_struct}
Time evolution of the position of the main maximum, $q_{\rm m}$, of the structure factor. The dashed black line indicates the deterministic linear prediction, $q_0$. Black circles (squares) correspond to predictions from Eq.\ (\ref{structure_factor_lin_fluids}) for $\sigma=10^{-2}$ ($\sigma=0$). Red up [blue down] triangles provide the position of $q_{\rm m, sto}(t)$ [$q_{\rm m,det}(t)$] as obtained in numerical simulations of Eq.\ (\ref{stochastic_thin_films}) for $\sigma=10^{-2}$ ($\sigma=0$). Rupture times are signalled by arrows. All results are obtained by averaging over $200$ noise realizations.
Inset: Number of droplets for different noise amplitudes at $t=220$. All lines are guides to the eye. }
\end{figure}

Although some spatial modulation can be seen for earlier times in Fig.\ \ref{fig:surface_in_time}, the system behavior is less visually clear.
However, at such times one can resort to linear stability analysis \cite{diez2007,grun2006,fetzer2007}. The time evolution of the system is conveniently described by the structure factor $S_q=\langle |h_q(t)|^2\rangle$, which within linear
approximation can be analytically obtained~\cite{grun2006,fetzer2007},
\begin{equation}
 S_q\,=\, (2\pi)^2 \left [ S_0(q) e^{2\omega(q)t}\,+\, \frac{\sigma^2 h_0^3}{2}\frac{q^2}{\omega(q)}\left(e^{2\omega(q)t}\,-\,1\right) \right].
 \label{structure_factor_lin_fluids}
\end{equation}
Here, $h_q(t)$ is the Fourier cosine transform \cite{boyd2001} of $h(x,t)$ for wavenumber $q$, $S_0(q)$ is the initial structure factor, $h_0$ is a film thickness, and the growth rate is given by the dispersion relation $\omega(q) \,=\,  h_0^3 q^2\left(2q_0^2-q^2\right)/3$. Here $q_0^2=-\Pi'(h_0)/2$. The wavelengths of unstable perturbations correspond to $q\in[0,\sqrt{2}q_0]$, for which $\omega(q) \geq 0$. Starting from an initial condition with mean $h_0$, the deterministic system very quickly selects the wavenumber $q_{\rm m,det}=q_0$ for which the growth rate $\omega(q)$ reaches its positive maximum, see black squares and blue triangles in Fig.\ \ref{fig:peak_struct}, where we plot the time evolution of the value of wavenumber $q_{\rm m}$ at which the main maximum of $S_q$ occurs.
For our parameter choice, $q_0=2.464$. Within linear approximation, this sets the length scale of the pattern, $\lambda=2\pi/q_{0}=2.546$, namely, the average size of surface undulations. In contrast, stochastic systems initially develop nontrivial short lengthscale (large $q$) correlations, so that $S_q(t)$ displays a maximum for a wavenumber value $q_{\rm m,sto}$ which decreases with time towards the deterministic value $q_0$, see Fig.\ \ref{fig:peak_struct} and \cite{SM_sq}.
This is the process described in \cite{fetzer2007} as coarsening.
Note that, as mentioned above, droplets have not yet formed; as seen in \cite{SM_sq}, for these times the film morphology remains largely a
small-amplitude sinusoid.  In addition, for stochastic simulations, $q_{\rm m,sto} >q_0$; as we will see, this inequality does not hold in the nonlinear regime.

\begin{figure}[t!]
 \vspace{0.5cm}
\includegraphics[width=8.5cm]{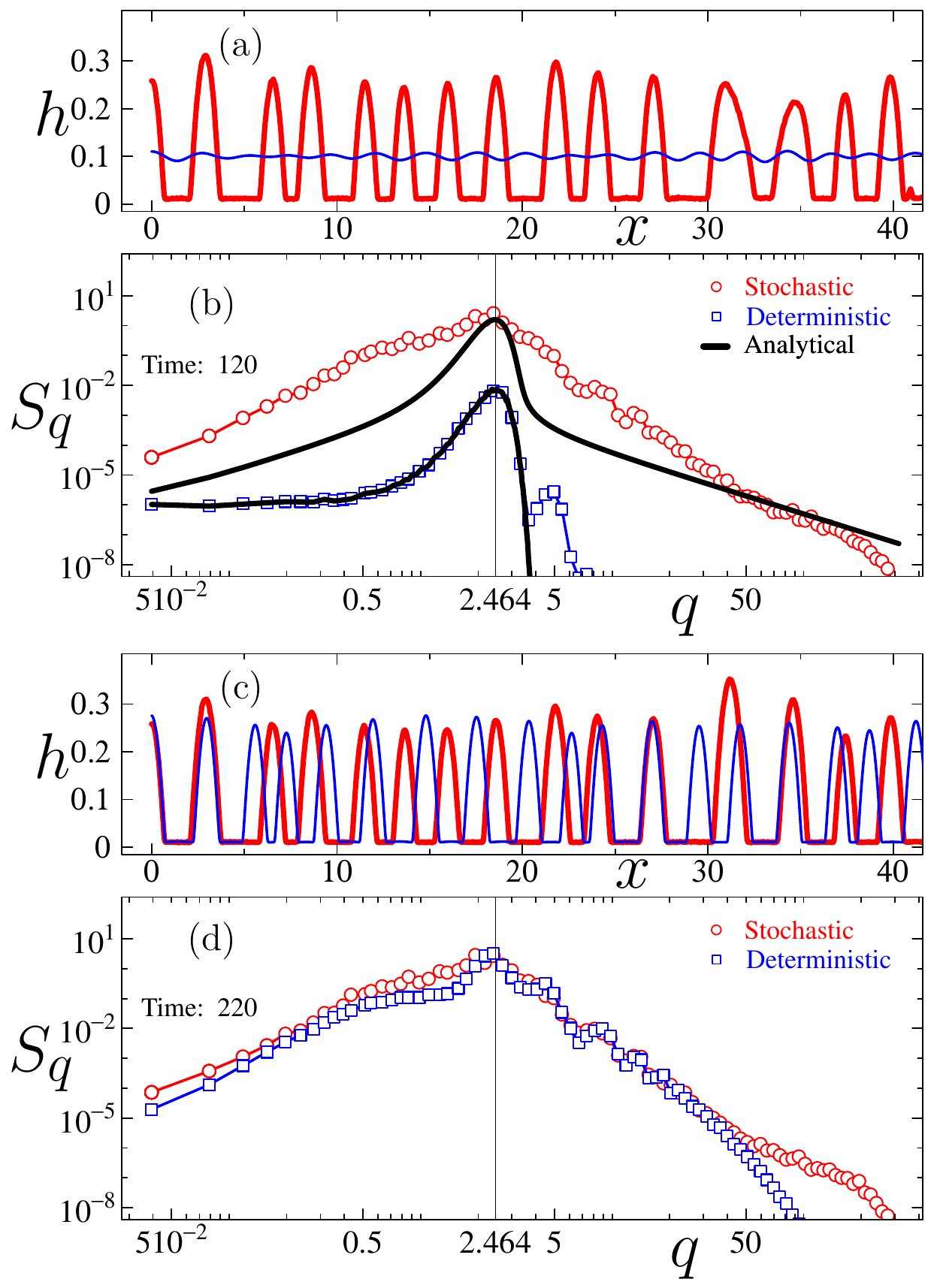}
\caption{\label{fig:shape_structure}
(a), (c): Surface morphologies from simulations of Eq.\ (\ref{stochastic_thin_films}) for $\sigma=0$ (blue line) and $\sigma=10^{-2}$ (single realization, thick red line) at $t=120$ (a) and $220$ (c). (b), (d): Structure factor averaged over $200$ noise realizations, at $t=120$ (b) and $220$ (d) for $\sigma=0$ (blue squares) and $\sigma=10^{-2}$ (red circles). The thick black lines in (b) provide the corresponding analytical predictions from Eq.\ (\ref{structure_factor_lin_fluids}). Thin lines are guides to the eye.}
\end{figure}

Within the range of validity of the linear approximation, the film develops unstable modes that remain independent of one another. If the linear predictions were applicable to long times, then the number of drops eventually formed would be essentially fixed by the linear value $\lambda=2.546$, since $S_q$ is characterized by a well defined peak around $q=q_{\rm m}$, see black lines in Fig.\ \ref{fig:shape_structure}.
However, experiments \cite{gonzalez2013} show that the distribution of droplet sizes is relatively wide. Droplet forms differ strongly from smooth sinusoids, and they interact non-trivially (e.g.\ through merging and coalescing) during their evolution. On long time scales, the number of drops needs to reduce, as a single larger drop is energetically more favorable than two smaller ones \cite{glasner2003,limary2003}. Actually, the most stable configuration of the system is a single droplet, since the evolution described by Eq.\ (\ref{stochastic_thin_films}) drives the system to the minimum of the interface Hamiltonian ${\cal H}[h] = \int {\rm d}x \, \left[\Phi(h)+\gamma (\partial_x h)^2/2\right]$ \cite{grun2006}.

Hence, we next need to address droplet formation for times $t\gtrsim 60$, see Fig. 2, away from the linear regime. As seen in the animation provided at \cite{SM_sq}, nonlinear effects indeed set in for $t\simeq 60$. Thus, the deterministic structure factor develops higher harmonics, while the stochastic $S_q$ also departs clearly from the linear solution, Eq.\ (\ref{structure_factor_lin_fluids}), see Fig.\ \ref{fig:shape_structure} for two sample times. The higher harmonics are at least one order of magnitude smaller than the main peak \cite{SM_sq}, so that they barely influence later stages of the evolution. In addition, the rupture time at which well-defined droplets form is much shorter for the stochastic ($t_{\rm r,sto}\simeq 80$) than in the deterministic ($t_{\rm r,det}\simeq 180$) case, see \cite{SM_sq} and also Fig.\ \ref{fig:shape_structure}(a) for $t=120$, where droplets have appeared in the former case, but not yet in the latter.

After rupture, the $S_q$ distribution broadens around the main peak both in the stochastic and in the deterministic systems, and for values of $q$ on both sides of $q_{\rm m}$, see Fig.\ \ref{fig:shape_structure}(b,d) and \cite{SM_sq}. Moreover, there is an additional boost in the rupture process so that stochastic droplets create faster than one would expect using the linear theory: Note that rupture times are signaled by a kink in the corresponding $q_{\rm m}(t)$ data. At rupture, nonlinear ripening of droplets takes place, namely, a decrease of $q_{\rm m}$ with time, which is more pronounced and occurs earlier in the stochastic system. In contrast to linear predictions, the deterministic system also undergoes a similar, albeit delayed process. We conjecture that disorder in droplet positions favors merging of nearby drops, inducing more rapid decrease of $q_{\rm m}$ in the stochastic system.

Also, for any $\sigma \geq 0$, once the drops are well formed the decrease in $q_{\rm m}(t)$ with time slows down. On average, the value of $q_{\rm m}$ which is eventually achieved (say, for $t\gtrsim 220$) is smaller for $\sigma\neq 0$. This behavior implies a smaller number of drops for a fixed domain in the stochastic system, see e.g.\ Fig.\ \ref{fig:shape_structure}(c). Also recall Fig.\ \ref{fig:surface_in_time}, where substantial drop merging is seen for this case during times from rupture up to $t\simeq120$. We note that, for the time scales considered so far, the final number of droplets decreases when the noise intensity (say, temperature) increases, see the inset of Fig.\ \ref{fig:peak_struct}.

\begin{figure}[t!]
\includegraphics[width=8.5cm]{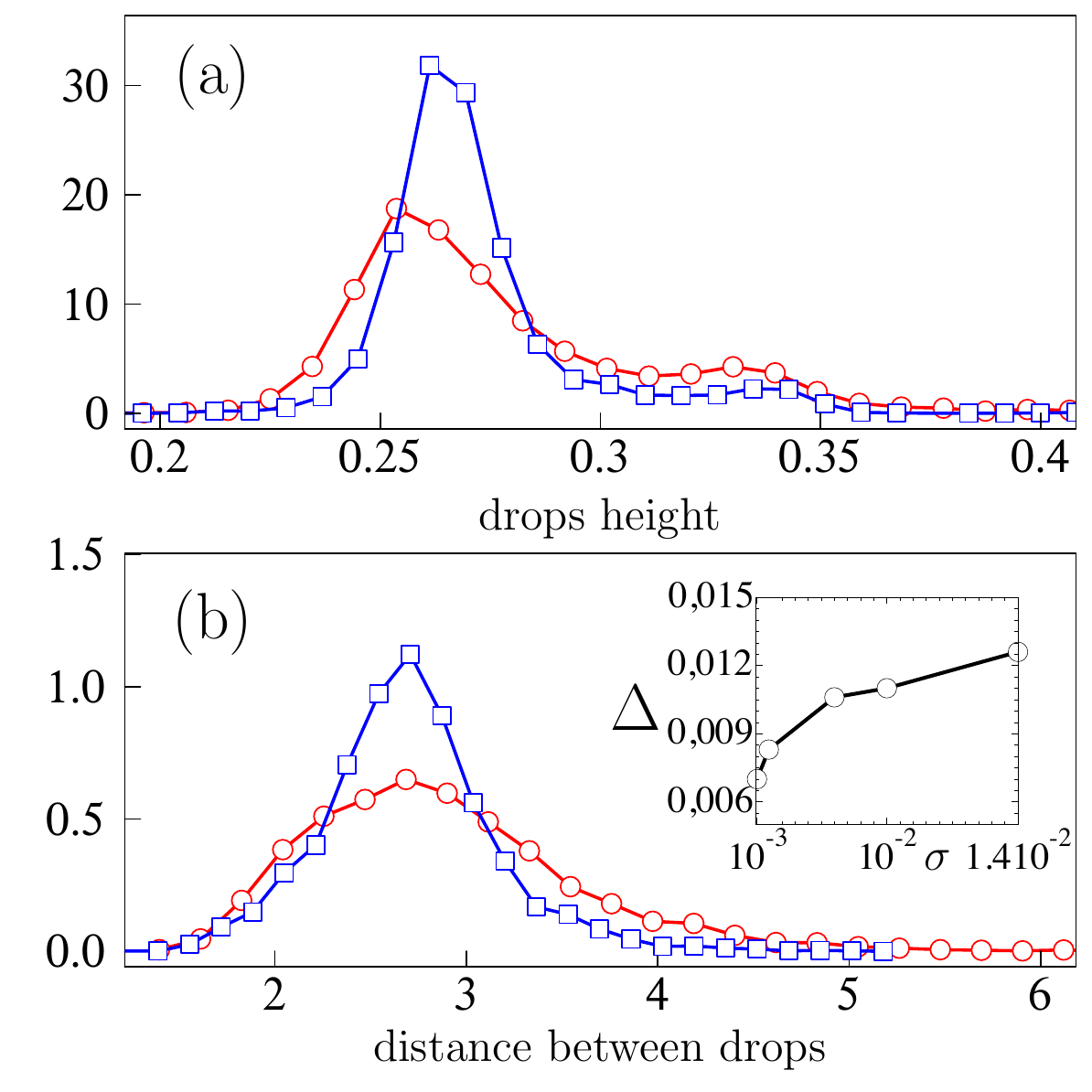}
\caption{\label{fig:histograms} Distribution functions of drop heights (a) and inter-drop distances (b) at time $t=220$ for $\sigma=0$ (blue squares) and $\sigma=10^{-2}$ (red circles). Inset: width of the main peak of $S_q$ at $t=220$, as a function of noise amplitude. All lines are guides to the eye.}
\end{figure}

Figure~\ref{fig:histograms}, showing the distribution of drop heights and their distances, illustrates a further significant difference between stochastic and deterministic evolution: stochasticity indeed leads to much wider droplet distributions, and therefore to much more irregular patterns. Two sample morphologies are
compared in Fig.\ \ref{fig:shape_structure}(c). Also, the inset in Fig.~\protect{\ref{fig:histograms}}(b) shows that the width at half maximum of the $S_q$ distribution, $\Delta$, is an increasing function of noise amplitude, as expected. This finding may be of significant importance in applications, where regularity of the distribution of drops is often desired. Our results suggest that decreasing noise amplitude may be the key to achieve this goal.

Finally, we consider much larger time scales, to probe the convergence of the stochastic system to the equilibrium single-drop solution. Up to the times discussed so
far, the decrease of $q_{\rm m}(t)$ seems mostly induced by droplet coalescence. This introduces relatively large distances among remaining units, recall Fig.\ \ref{fig:surface_in_time} for long times. For still longer times, droplet interaction occurs mostly through the precursor layer, inducing non-interrupted coarsening of the pattern into a single drop morphology \cite{glasner2003}. For well-separated droplets and $\sigma=0$, analytical predictions actually exist for the decrease of the number of droplets $N(t)$ with time \cite{glasner2003}. We have considered the evolution predicted by Eq.\ (\ref{stochastic_thin_films}) for $\sigma\neq 0$ at very long times, up to $t=10^4$. Computational feasibility requires a larger precursor thickness $h_*=0.04$ and smaller noise, $\sigma=10^{-5/2}$. Our results indicate that fluctuations do shorten significantly the time scales on which coarsening occurs. However, they become less relevant with increasing time, to the extent that the asymptotic behavior of $N(t)$ is not modified with respect to the deterministic case, at least within the accuracy of the results, see \cite{SM_long_t}. Thus, droplet coarsening counts among phenomena for which noise does not change the coarsening universality class \cite{bray1994} of the corresponding deterministic system.

In summary, we have shown that stochastic effects due to thermal noise may play a significant role in dewetting of thin fluid films, in each of the three stages of evolution
considered. For very early times, stochasticity leads to a decrease of the most unstable wavenumber, $q_{\rm m,sto}$, down from the values that are large compared to the deterministic one, $q_0$; however, within this stage $q_{\rm m,sto}$ remains larger than $q_0$. After this, noise triggers an earlier onset of nonlinear effects, inducing a shorter rupture time. At these time scales, stochasticity leads to droplet coarsening, in the sense that $q_{\rm m,sto}<q_0$, in contrast to the linear regime. Finally, for much longer times, fluctuations speed up the coarsening process that will ultimately lead to formation of the single-drop, energetically favored state. Qualitatively, the deterministic coarsening law for the number of drops remains unchanged. However, {\em quantitatively} the time scales involved in this long-time coarsening process are significantly influenced by noise, and we conjecture that stochastic effects may be observable in careful experiments carried out with fluid films of nanoscale thickness.

Partial support for this work has been provided by MINECO (Spain) grants No.\ FIS2010-22047-C05-04 and No.\ FIS2012-38866-C05-01,
and by NSF (USA) grant No. CBET-1235710.  S.\ N.\ acknowledges support by Universidad Carlos III de Madrid.

\newpage

\onecolumngrid

\begin{center}
\large \textbf{Supplemental Material}
\end{center}

In this Supplemental Material we provide results from numerical simulations of the stochastic thin film equation discussed in the main text, which in the corresponding (hatted) dimensionless units reads
\begin{equation}
  \partial_t h\;=\;  \nabla \cdot \left\{h^3\,\nabla\left[- \nabla^2 h - \Pi(h) \right]+ \sigma\, h^{3/2}\,\epsilon(x,t)\right\},
 \label{stohastic_thin_films}
\end{equation}
where $\epsilon$ is an uncorrelated Gaussian noise with zero mean and unit variance, and the hats have been removed to simplify the notation.

The film morphology $h(x,t)$, the structure factor $S_q(t)$, together with the histograms of individual drop heights and inter-drop distances, are all shown in Movie S1 for times $t\in[1,250]$, which includes deterministic ($\sigma=0$) and stochastic ($\sigma=10^{-2}$) simulations of Eq.\ (\ref{stohastic_thin_films}). These simulations have been performed using as initial condition a film of non-dimensional height $h_0=0.1$ (perturbed by Gaussian white noise with $10^{-3}$ amplitude), with a precursor thickness $h_*=0.01$. Other parameters are described in the main text.

\begin{figure}[h]
\includegraphics[width=12cm]{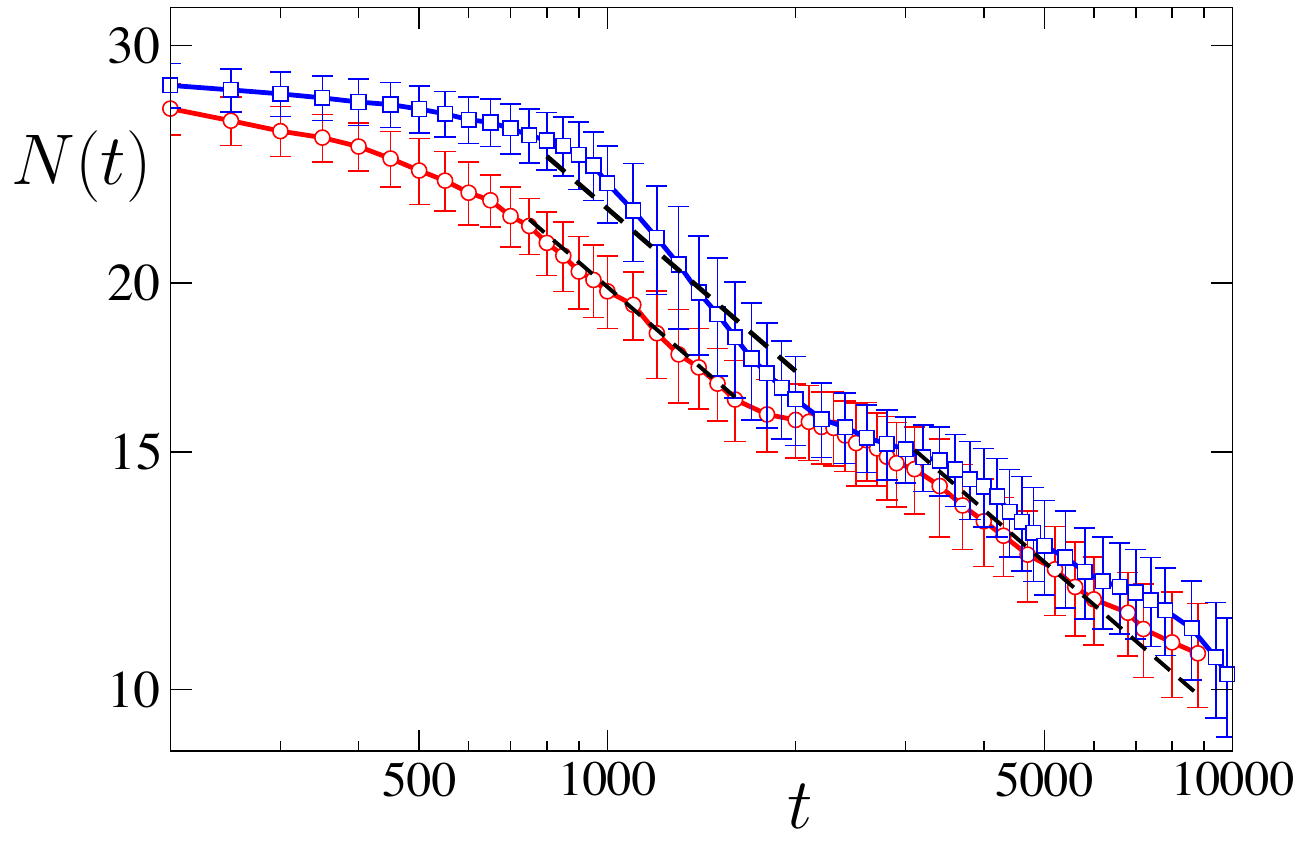}
\caption{Log-log plot showing the time evolution of the number of droplets vs time, for long times up to $t=10000$. Blue squares (red circles) correspond to deterministic ($\sigma=0$) and stochastic ($\sigma=10^{5/2}$) simulations, respectively. For reference, the dashed lines correspond to the power-law decay $N(t) \sim t^{-2/5}$ found for very large deterministic systems in \cite{glasner2003}. Solid blue and red lines are guides to the eye.}
\label{fig:witelski}
\end{figure}

In order to probe the long-time evolution of the droplet pattern, in which coarsening takes place mainly through the precursor film that communicates individual droplets \cite{glasner2003}, we have considered a larger precursor thickness, $h_*=0.04$. Computational feasibility has also required us to set $\sigma=10^{-5/2}$. The time evolution of the film morphology under these conditions can be assessed in Movie S2, which corresponds to stochastic simulations. Figure \ref{fig:witelski} is a plot of the number of droplets vs time, $N(t)$, in this long-time coarsening regime, averaged over 40 realizations of the noise. While the decay of $N(t)$ may not be far from analytical predictions and simulations for very large discrete models \cite{glasner2003}, our simulations seem to support the irrelevance of noise at sufficiently long times $t\gtrsim 2000$.
\vspace{0.3cm}

\end{document}